\documentstyle[preprint,prl,aps,psfig]{revtex}
\begin{document}
\draft


\title{Electron Transport in Mesoscopic Disordered SNS Junctions}  

\author{Athanassios Bardas and Dmitri V. Averin } 

\address{Department of Physics,\\
 State University of New York at Stony Brook,\\
 Stony Brook NY 11794-3800}

\date{\today}
\maketitle
\begin{abstract}

We have generalized the scattering--matrix theory of multiple 
Andreev reflections in mesoscopic Josephson junctions to the 
multi-mode case, and applied it to short 
superconductor/normal metal/superconductor junctions with diffusive 
electron transport. Both the {\em dc} and {\em ac} current-voltage 
characteristics are analyzed for a wide range of bias voltages $V$.  
For voltages smaller than the supeconducting gap the dc differential 
conductance of the junction diverges as $1/\sqrt{V}$.

\end{abstract}


Considerable interest, both in experiments \cite{kl,van,fr,mur,ku,sac} 
and in theory \cite{sh1,we1,gun}, is attracted currently 
to finite-voltage transport properties of mesoscopic Josephson junctions 
with high electron transparency. The mechanism of conduction in such 
junctions is the process of multiple Andreev reflections (MAR) 
\cite{kbt,zai1} that takes place at the interfaces of the junction 
scattering region, and bulk superconducting electrodes. 
Phenomelogically, the MAR processes manifest themselves in the so-called 
``subharmonic gap structure'' (SGS) in the current-voltage ($IV$) 
characteristics of the junction: current
singularities at voltages $V_n=2\Delta/en$, $n=1,2,...$, where $\Delta$
is the superconducting energy gap of the junction electrodes. 
Recently developed quantitative description of MAR in junctions with 
arbitrary electron transparency $D$ \cite{we1} is based on the 
usual scattering approach for Bogolyubov-de Gennes equations 
\cite{fur,been,bag}, combined 
with the idea of acceleration of quasiparticles in the junction by 
non-vanishing bias voltage. Point contacts fabricated with the 
controllable break junction technique \cite{van,sac,oub} allow for 
accurate 
comparison between theory and experiment. In particular, the most 
recent experiment \cite{sac} found that the $I-V$ characteristics of 
aluminum point contacts can be well explained by theory 
for the whole range of the contact transparencies $D$, from 
the tunnel junction limit $D\rightarrow 0$ all the way to the 
ballistic contacts with $D\rightarrow 1$. 

So far the scattering theory of MAR has been formulated only for 
junctions with one propagating electron mode. The aim of this 
work is to extend the theory to the multi-mode case and apply it to 
diffusive SNS junctions. We show that if the scattering matrix $S$ 
of the junction does not depend on energy on the scale of $\Delta$, 
the time-dependent current in the junction can be represented as a 
sum of independent contributions from individual transverse modes. 
Therefore, the current depends only on the distribution of the 
transmission probabilities $D_m$ of these modes. 

We assume that the junction length $L$ (Fig.\ 1) is smaller than the 
coherence length $\xi$ of the superconducting electrodes of the 
junction and inelastic scattering length $l_{in}$:\,  $L\ll \xi,\, 
l_{in}$. In this case the transport properties of the junction are 
determined by the interplay of scattering inside the
 junction characterized by the scattering 
matrix $S$, and the Andreev reflection with amplitude $a(\varepsilon)$ 
at the two interfaces between the junction scattering region and 
bulk superconducting electrodes which are in equilibrium,  
\begin{equation}
a(\varepsilon)=\frac{1}{\Delta} \left\{ \begin{array}{ll} 
\varepsilon- \mbox{sign}(\varepsilon) (\varepsilon^2- \Delta^2)^{1/2} 
\, , \;\;\; & \mid \varepsilon \mid > \Delta\, , \\
\varepsilon-i(\Delta^2 -\varepsilon^2)^{1/2}\, , \;\;\; & \mid 
\varepsilon \mid < \Delta\, . \end{array} \right. 
\label{aa} \end{equation}  
Here $\varepsilon$ is the quasiparticle energy relative to the Fermi 
level of the electrode. The scattering matrix $S$ is a unitary and 
symmetric matrix $2N \times 2N$, where $N$ is the number of propagating 
transverse modes supported by the junction, and can be written in terms 
of reflection and transmission $N\times N$ matrices $r$, $t$:  
\begin{equation} 
S= \left( \begin{array}{cc} r & t \\ t' & r'
\end{array} \right) \, ,
\label{c5}  \end{equation} 
where $t'=t^T$, $r'= -(t^*)^{-1}r^{\dagger} t$, and $tt^{\dagger} +
rr^{\dagger} =1$.   

Due to acceleration of quasiparticles by the 
applied bias voltage, an electron with energy $\varepsilon$ emerging 
from one of the electrodes generates electron and hole states at 
energies $\varepsilon +2neV$ with arbitrary $n$. Thus, the electron 
and hole wavefunctions in regions I and II of the junction  
(Fig.\ 1) can be written as follows: 
\begin{equation} 
\mbox{(I)} \;\;\;\; \begin{array}{l} 
\psi_{el} =\sum_{n}[(a_{2n}A_n+J\delta_{n0})e^{ikx}+B_ne^{-ikx} ] 
 e^{-i(\varepsilon +2neV)t/\hbar }\, ,  \\ 
\psi_{h} =\sum_{n}[A_ne^{ikx}+a_{2n}B_ne^{-ikx} ] 
e^{-i(\epsilon +2neV)t/\hbar } \, , \end{array} 
\label{I} \end{equation}
\begin{equation} 
\mbox{(II)} \;\;\;\; \begin{array}{l} 
\psi_{el} =\sum_{n}[C_ne^{ikx}+a_{2n+1} D_ne^{-ikx} ] 
 e^{-i(\varepsilon +(2n+1)eV)t/\hbar }\, ,  \\ 
\psi_{h} =\sum_{n}[a_{2n+1}C_ne^{ikx}+D_ne^{-ikx} ] 
e^{-i(\varepsilon +(2n+1)eV)t/\hbar } \, , \end{array} 
\label{II} \end{equation}
where $a_n\equiv a(\varepsilon+neV)$. In these equations we took 
into account that the amplitudes of electron and hole wavefunctions 
are related via Andreev reflection. Furthermore, we neglected variations 
of the quasiparticle momentum $k$ with energy assuming that the 
Fermi energy in the electrodes is much larger than $\Delta$. The 
quasiparticle energies in regions I and II are measured relative to 
the Fermi level in the left and right electrode, respectively. 

The amplitudes of electron and hole 
wavefunctions have transverse mode index $m$ not shown in eqs.\ 
(\ref{I}) and (\ref{II}), e.g., $A_n\equiv \{ A_{n,m} \}$, $m=1,...,N$. 
The source term $J$ describes an electron generated in the $j$th 
transverse mode by a quasiparticle incident on the constriction from 
the left superconductor: $J(\varepsilon)=(1-\mid \! a_0 \! \mid ^2)^{1/2} 
\delta_{mj}$. The current $I(t)$ in the constriction can be calculated 
in terms of the wavefunction amplitudes. Equation (\ref{I}) and 
(\ref{II}) imply that the current oscillates with the Josephson 
frequency $\omega_J =2eV/\hbar$ and can be expanded in Fourier 
components: $I(t)=\sum_k I_k e^{ik\omega_J}$. Summing the 
contributions to the current from electrons incident both 
from the left and right superconductors at different energies 
$\varepsilon$ we obtain the Fourier components $I_k$: 
\[ I_k =-\frac{e}{\pi \hbar} \int_{-\mu-eV}^{\mu} d\epsilon \tanh 
\{ \frac{\epsilon}{2T} \} \mbox{Tr}[ (J J^{\dagger} \delta_{k0} + 
a_{2k}^* JA_k^{\dagger} + a_{-2k} A_{-k} J^{\dagger} + \] 
\begin{equation} 
\left.  + \sum_n (1+a_{2n}a_{2(n+k)}^*) (A_nA_{n+k}^{\dagger}  
-B_nB_{n+k}^{\dagger} )) ] \right|_{\mu \rightarrow \infty} \, ,  
\label{c2} \end{equation}   
where Tr is taken over the transverse modes.

Amplitudes of the wavefunctions (\ref{I}) and (\ref{II}) are related 
via the scattering matrix $S$. Taking into account that the scattering 
matrix for the holes is the time-reversal of the electron 
scattering matrix $S$, we can write:   
\begin{equation} 
\left( \begin{array}{c} B_n \\ C_n \end{array} \right)  = S 
\left( \begin{array}{c} a_{2n}A_n +J\delta_{n0} \\ 
a_{2n+1}D_n \end{array} \right) \, , 
\label{c3} \end{equation} 
\begin{equation} 
\left( \begin{array}{c} A_n \\ D_{n-1} \end{array} \right) = 
S^* \left( \begin{array}{c} a_{2n}B_n \\ 
a_{2n-1}C_{n-1} \end{array} \right) \, , 
\label{c4}  \end{equation} 
Eliminating $A_n$ between eq.\ (\ref{c3}) and inverse of  
eq.\ (\ref{c4}), we find a relation between the amplitudes 
$B_n$ and $D_n$. Combining this relation with the expression for $D_n$ 
in terms of $B_n$ that follows from the inverse of eq.\ (\ref{c3}) 
and eq.\ (\ref{c4}) we arrive at the following recurrence relation 
for $B_n$: 
\[ tt^{\dagger} \left( \frac{a_{2n+2}a_{2n+1}}{1-a_{2n+1}^{2}} B_{n+1} -
(\frac{a_{2n+1}^2}{1-a_{2n+1}^{2}} + \frac{a_{2n}^2}{1-a_{2n-1}
^{2}}) B_n +   \frac{a_{2n}a_{2n-1}}{1-a_{2n-1}^{2}} B_{n-1} \right) - \] 
\begin{equation} 
- [1-a_{2n}^{2}] B_n = - rJ \delta_{n0} \, . 
\label{c6} \end{equation} 
Since the hermitian matrix $tt^{\dagger}$ can always be diagonalized 
by an appropriate unitary transformation $U$, the recurrence relation 
(\ref{c6}) implies that the structure of the amplitudes 
$B_n$ as vectors in the transverse-mode space is:
\begin{equation} 
B_n = U^{\dagger} f_n(D) U r J \, ,
\label{c8} \end{equation} 
where $D=U tt^{\dagger} U^{\dagger}$ is the diagonal matrix of 
transmission probabilities $D_m$, $m=1,...,N$. The functions $f_n(D)$ 
are determined by the solution of the recurrence relation (\ref{c6}) 
with the diagonalized transmission matrix $tt^{\dagger}$.   
	
Equation (\ref{c8}) shows that the contribution of the amplitudes 
$B_n$ to the currents (\ref{c2}) can be written as 
\[ \mbox{Tr} [B_n B_{n'}^{\dagger} ] = (1-\mid \! a_0 \! \mid ^2) 
\mbox{Tr}  [ f_n(D) f_{n'}^*(D) (1-D) ] \, , \] 
i.e., it can be represented as a sum of independent contributions 
from different transverse modes with the transparencies $D_m$. 
Following similar steps we can show that the same is true for the 
amplitudes $A_n$. Therefore, the Fourier components (\ref{c2}) of 
the total current can be written as sums of independent contributions 
from individual transverse modes: 
\begin{equation} 
I_k=\sum_m I_k(D_m) \, , 
\label{cc9} \end{equation} 
where the contribution of one (spin-degenerate) mode is:    
\[ I_k(D) =\frac{e}{\pi \hbar} \left[ eV D\delta_{k0} - \int d\epsilon 
\tanh \{ \frac{\epsilon}{2T} \} (1-|a_0|^2)(a_{2k}^*A_k^* +a_{-2k} 
A_{-k} \right. \] 
\begin{equation} 
\left. + \sum_n (1+a_{2n}a_{2(n+k)}^*) 
(A_n A_{n+k}^* -B_n B_{n+k}^*)) \right]  \, .  
\label{ccc9} \end{equation}
with the integral over $\varepsilon$ taken in large symmetric limits. 
The amplitudes $B_n$ in this equation are determined by the 
recurrence relation which follows directly from eq.\ (\ref{c6}):   
\[ D\frac{a_{2n+2}a_{2n+1}}{1-a_{2n+1}^{2}} B_{n+1} - 
[D(\frac{a_{2n+1}^2}{1-a_{2n+1}^{2}} + \frac{a_{2n}^2}{1-a_{2n-1}
^{2}})+1-a_{2n}^{2}] B_n + \] 
\begin{equation}
+ D\frac{a_{2n}a_{2n-1}}{1-a_{2n-1}^{2}} B_{n-1} = - R^{1/2}\delta_{n0} 
\, ,  \;\;\; R\equiv 1-D\, . 
\label{c9} \end{equation}
Instead of using similar independent recurrence relation for $A_n$, it 
is more convenient to determine these coefficients from an equivalent 
relation that can be obtained from a single-mode version of eqs.\ (\ref{c3}) 
and (\ref{c4}) \cite{we1}:  
\begin{equation}  
A_{n+1}-a_{2n+1}a_{2n}A_n =R^{1/2}(B_{n+1}a_{2n+2}-B_na_{2n+1})  
+a_1\delta_{n0} \, ,  
\label{c10} \end{equation} 

Equations (\ref{cc9}) -- (\ref{c10}) completely determine the 
time-dependent current in a short constriction with arbitrary 
distribution of transmission probabilities. In particular, we can use 
them to calculate the current in a short disordered SNS 
junction with large number of transverse modes $N\gg 1$ and diffusive 
electron transport in the N region. In this case, the distribution of 
transmission probabilities is quasicontinuous, and is characterized 
by the density function $\rho(D)$ (see, e.g., \cite{naz} and references 
therein): 
\begin{equation}  
\sum _m ... = \int_0^1 dD \rho(D) ... \, , \;\;\;\; \rho(D) = \frac{\pi 
\hbar G}{2e^2} \frac{1}{D(1-D)^{1/2} } \, , 
\label{c11} \end{equation} 
where $G$ is the normal-state conductance of the N region.

Figure 2 shows the results of the numerical calculations of the dc 
current-voltage ($IV$) characteristics and differential conductance of 
the short SNS Josephson junction based on eqs.\ (\ref{ccc9}) -- (\ref{c11}). 
We see that the $IV$ 
characteristics have all qualitative features of highly transparent
Josephson junctions: subgap current singularities at $eV=2\Delta/n$ and 
excess current $I_{ex}$ at $eV\gg 2\Delta$. It is instructive to compare 
quantitatively these features to those in  the $IV$ characteristics of a 
single-mode Josephson junctions plotted in \cite{we1}. Such a comparison 
shows that the magnitude of the excess current in the SNS junction, as well  
as the overall level of current in the sub-gap region correspond 
approximately 
to a single-mode junction with large transparency $D\simeq 0.8$. At the 
same time, the subharmonic gap structure and the gap feature at 
$eV=2\Delta$ are much more pronounced than in a single-mode junction of 
this transparency. The amplitude of the oscillations of the differential 
conductance corresponds roughly to the junction with $D\simeq 0.4$ (although  
this comparison is not very accurate because of the different shapes of 
the curves). This ``discrepancy'' reflects the two-peak structure of the 
transparency distribution (\ref{c11}) of the diffusive conductor: the 
abundance of nearly ballistic modes leads to large excess and subgap 
currents, while the peak at low transparencies determines the SGS 
features. 

At large and small bias voltages the time-dependent current 
through the constriction $I(\varphi)$, where $\varphi = 2eVt/\hbar$, 
can be found analytically. At large voltages, $eV\gg \Delta$, the 
probability of MAR cycles decreases rapidly with the number of Andreev 
reflections in them. This implies that the higher-order harmonics 
$I_k$ of the current decrease rapidly with increasing $k$, and 
we can limit ourselves to the first harmonic: 
\[ I(\varphi) =I(V) + 2\mbox{Re}{I_1} \cos (\varphi) +   
2\mbox{Im}{I_1} \sin (\varphi) \, \] 
Truncating then the recurrence relations (\ref{c9}) and 
(\ref{c10}) at the coefficients $B_{\pm 1}$ and $A_{0,1}$ we can solve them 
explicitly and find the current from eq.\ (\ref{ccc9}). For a single 
mode at $T\ll \Delta$ we get: 
\begin{eqnarray} 
I(V) & = & \frac{eD}{\pi \hbar} \left[ eV+ \frac{\Delta D}{R} (1-
\frac{D^2}{2\sqrt{R}(1+R)} \ln (\frac{1+\sqrt{R}}{1-\sqrt{R}}) ) 
-\frac{\Delta^2}{2eV} \right] \, , \nonumber \\
\mbox{Im}{I_1} (V) & =  & \frac{\Delta^2}{\hbar V}\frac{DR}{1+R} \, , 
\label{c12} \\ 
\mbox{Re}{I_1} (V) & = &  -\frac{D \Delta^2}{\pi \hbar V} [R \ln{ 
\frac{eV}{\Delta}} + \frac{1+R}{2} \ln{2} + \frac{D}{2}(1+
\frac{1+R}{R}\ln{D} ) ] \, . \nonumber 
\end{eqnarray}
Expression for the excess current (second term in $I(V)$) 
was obtained before in \cite{mr,sh2}. Asymptotics of the ac components 
of the current agree both with the known results for the tunnel junction 
limit ($D\rightarrow 0$) and ballistic junction ($D\rightarrow 1$) 
\cite{we1}.

Averaging eqs.\ (\ref{c12}) with the distribution (\ref{c11}) we obtain 
the large-voltage asymptote of the current in the SNS junction: 
\begin{equation} 
I(V) = G [V+\frac{\Delta}{e} (\frac{\pi^2}{4}-1) -
\frac{\Delta^2}{2eV}) ] \, , 
\label{c13} \end{equation} 

\vspace*{-1ex}  

\[ \mbox{Im}{I_1} (V) =  \frac{\pi G \Delta^2}{e^2V}(1-\frac{\pi}{4}) 
\, , \;\;\;\;\;  
\mbox{Re}{I_1} (V) =  -\frac{G\Delta^2}{3e^2V} [\ln{(\frac{eV}
{4\Delta})}+ \frac{7}{3} ] \, . \] 
The second term in equation for $I(V)$ represents the excess current 
and was first found in \cite{avz} by the quasiclassical Green's function 
method. It can be checked that the asymptote (\ref{c13}) agrees well 
with the numerically calculated zero-temperature $IV$ characteristic 
shown in Fig.\ 2a. 

Analytical results at low voltages, $eV\ll \Delta$ can be obtained 
using the understanding\cite{we1} that the small voltage $V$ drives the 
Landau-Zener transitions between the Andreev-bound states of the modes 
with small reflection coefficients $R\ll 1$. Averaging the nonequilibrium 
voltage-induced contribution to the current (eq.\ (11) of Ref.\ 
\cite{we1}) with the distribution (\ref{c11}) we find the nonequilibrium 
part of the dynamic current-phase relation of a short SNS junction at 
$eV\ll \Delta$:  
\begin{equation}
I(\varphi) = \frac{\pi G\Delta \eta(\Delta)}{e} \left\{ \begin{array}{ll}  
0\, ,  & 0< \varphi <\pi \, , \\ 
 \sqrt{eV/\Delta} \sin(\varphi/2)  \, , & \pi < \varphi <2\pi \, ,  
\end{array} \right.    
\label{c15}\end{equation} 
where $\eta(\Delta)\equiv \tanh{(\Delta/2T)}$.
(Equilibrium part of the current-phase relation has been found before 
in \cite{ko2}.) From this equation we can find the voltage dependence 
of the Fourier harmonics of the time-dependent current at low voltages. 
The amplitudes of the first harmonics calculated numerically for 
arbitrary voltages are shown in Fig.\ 3. The curves agree with the 
high-voltage (\ref{c12}) and low-voltage (\ref{c15}) asymptotics and  
exhibit the SGS singularities at intermediate voltages. In general, the 
curves look qualitatively similar to those for the single-mode junction 
with intermediate transparency $D$ (see Figs.\ 2b,c  in Ref.\ \cite{we1}). 

Equation (\ref{c15}) implies that the dc differential conductance of 
the junction has the square-root singularity at $V\rightarrow 0$. 
\begin{equation} 
G(V) =\frac{dI}{dV} = \frac{\eta(\Delta)}{2} G \sqrt{\Delta/eV} \, . 
\label{c16} \end{equation} 
This singularity reflects directly the high-transparency peak in the 
distribution (\ref{c11}) and is not unique to the diffusive SNS junctions. 
A junction with the strongly disordered tunnel barrier \cite{bau} should 
exhibit the same $1/\sqrt{V}$ 
singularity with the prefactor $1/2$ in eq.\ (\ref{c16}) replaced 
with $2/\pi$. Physically, the origin of this conductance singularity is 
overheating of electrons in the junction by MAR. Electrons with energies 
inside the energy gap traverse the junction many times and as a result 
are accelerated to energies much larger the $eV$. This means that the 
effective voltage drop across the junction is much larger than $V$, 
leading to increased conductance. This mechanism of conductance 
enhancement is qualitatively similar to the so-called ``stimulation of 
superconductivity''\cite{al} (which is one of the plausible 
explanations of the zero-bias conductance singularities \cite{kr} 
in long semiconductor Josephson junctions), although quantitatively 
the phenomena are quite different. The fact that 
the singularity (\ref{c16}) is caused by electron overheating implies 
that at very low voltages it should be regularized by any mechanism of 
inelastic scattering. Nevertheless, in junctions shorter than inelastic 
scattering length $l_{in}$, there should be a voltage range where 
the conductance follows the $1/\sqrt{V}$ behavior. 
  
In summary, we have developed a theory of multiple Andreev reflections 
in multi-mode Josephson junctions and applied it to the diffusive 
SNS junctions and disordered tunnel barriers. The hallmark of the MAR 
processes in these systems is the zero-bias $1/\sqrt{V}$ singularity of 
the dc differential conductance. We have also calculated the low- and 
high-voltage asymptotes of the ac components of the time-dependent 
current in the SNS junctions.   

We would like to thank K. Likharev for fruitful discussions. 
This work was supported by DOD URI through AFOSR Grant 
\# F49620-95-I-0415 and by ONR grant \# N00014-95-1-0762.

\vspace{-3in}
\begin{figure}
\centerline{
\psfig{file=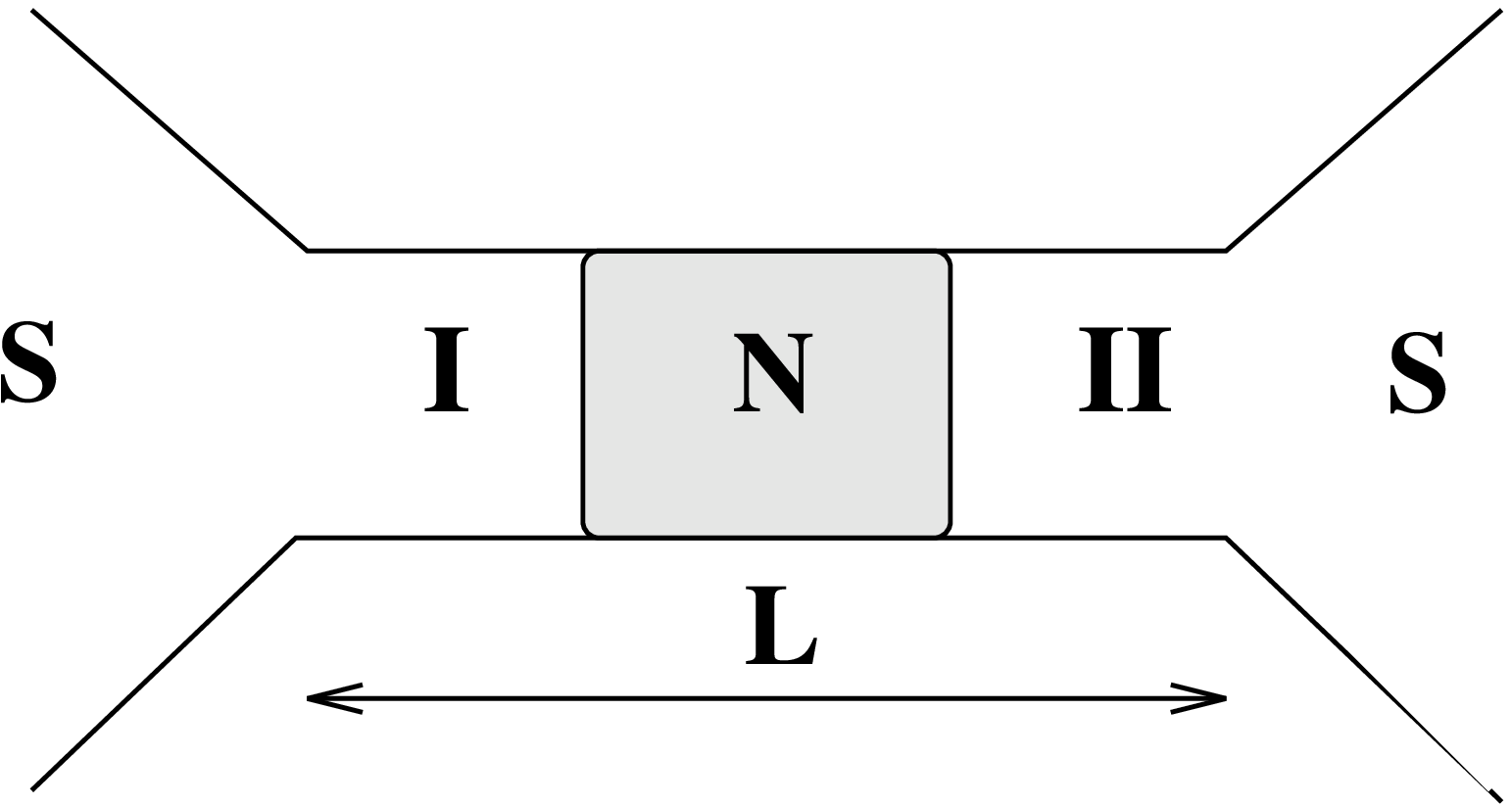,height=3.5in,width=4.5in,angle=0}}
\caption{
Schematic diagram of the mesoscopic disordered 
Josephson junction. I and II denote the portions of the
contact region separated by the scattering region (hatched) 
where the motion of the quasiparticles is diffusive. 
}\label{f1} 
\end{figure}

\newpage

\begin{figure}
\centerline{
\psfig{file=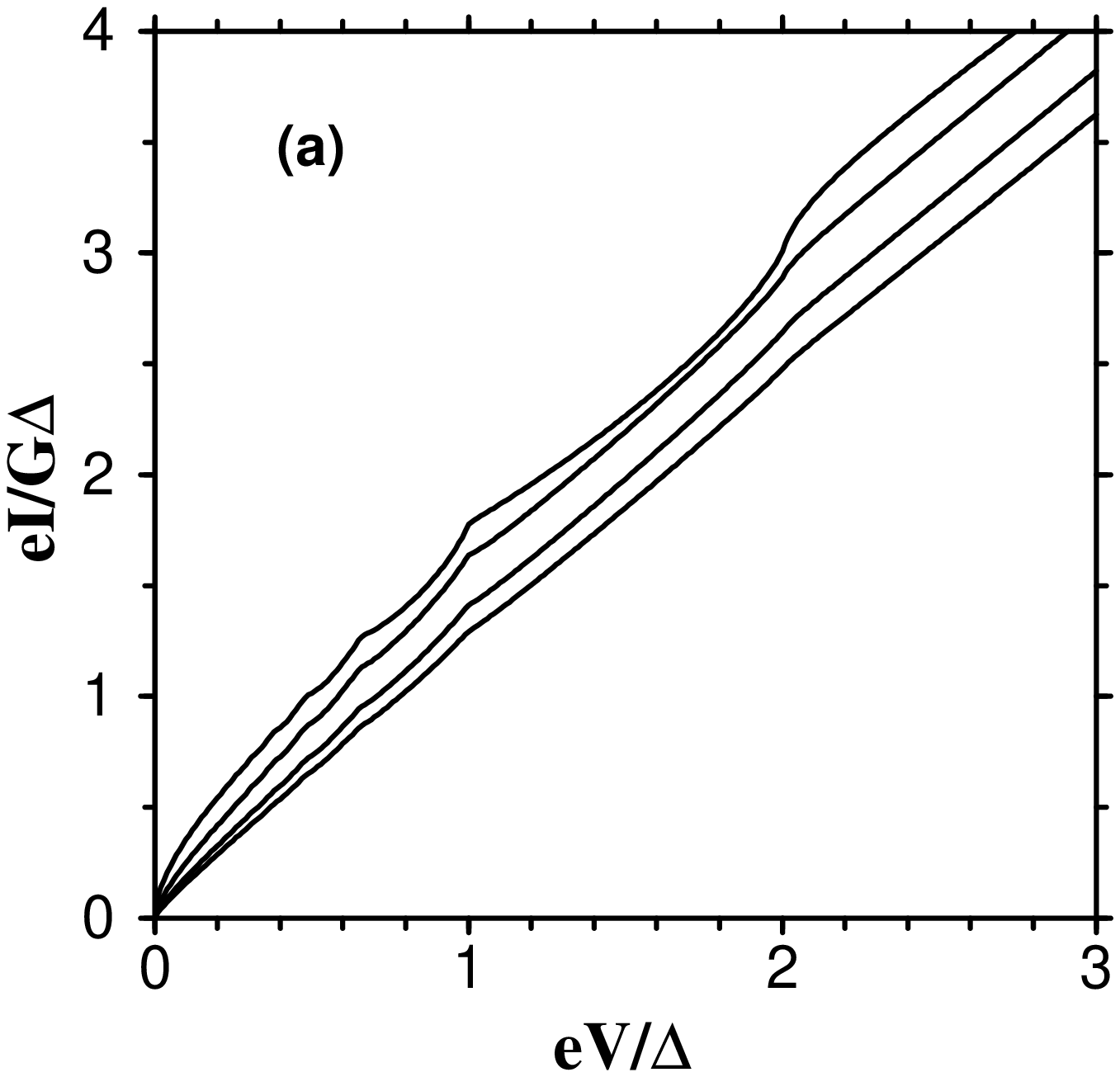,height=3.5in,angle=0}}
\centerline{
\psfig{file=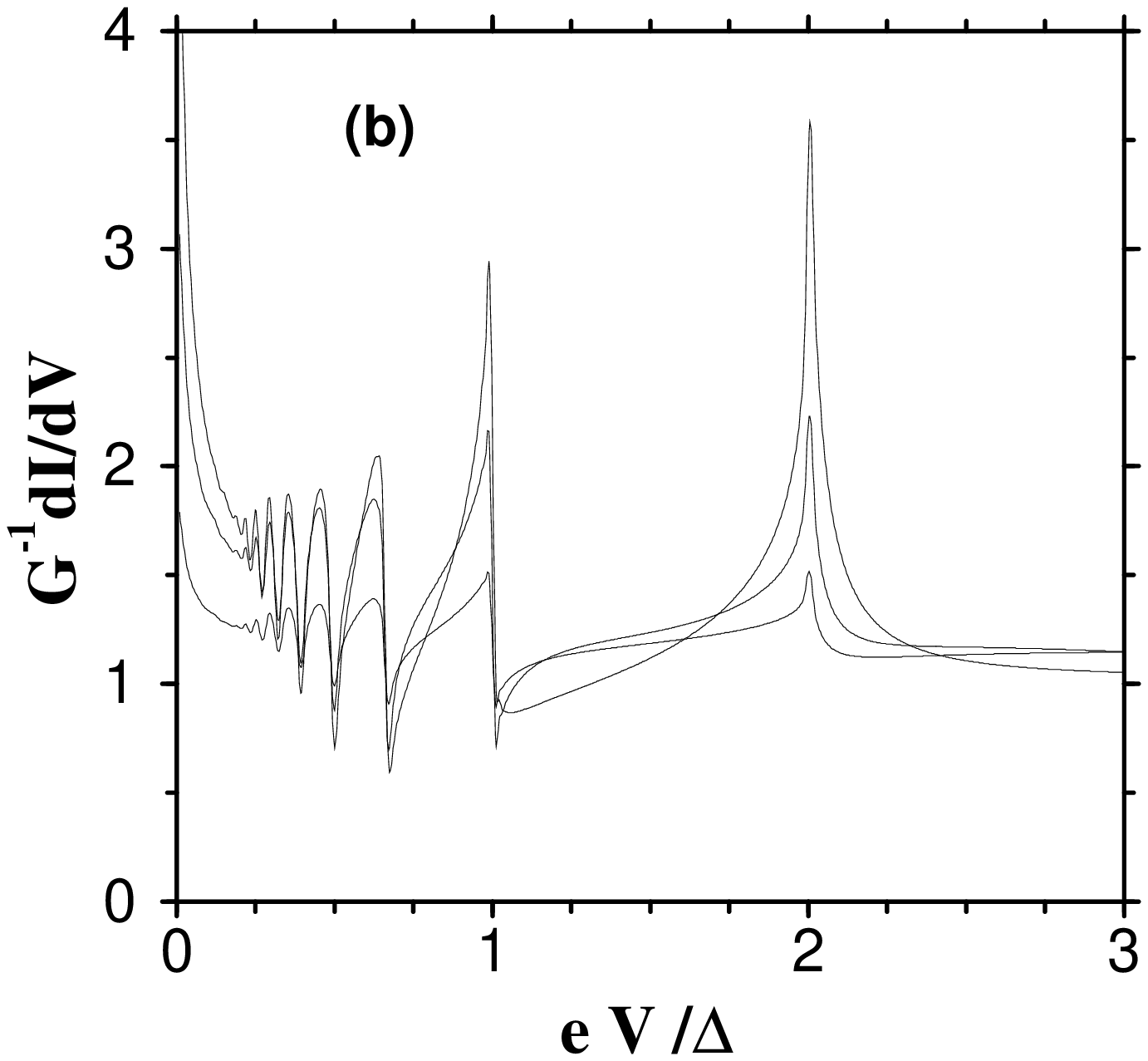,height=3.5in,angle=0}
}
\caption{
The quasiparticle current (a) and the
differential conductance (b) versus voltage at 
various temperatures, for a disordered SNS junction. 
From top to bottom $T= 0, 1, 2, 3 \Delta$. 
At low voltages, the dc current (a) has a square root dependence 
on voltage, while at high voltages exhibits excess current. 
The conductance (b) posseses subharmonic singularities, 
diverges at low voltages, and at high voltages asympotically
tends to the normal state conductance of the disordered region.
}   
\label{f2} 
\end{figure}

\begin{figure}
\centerline{
\psfig{file=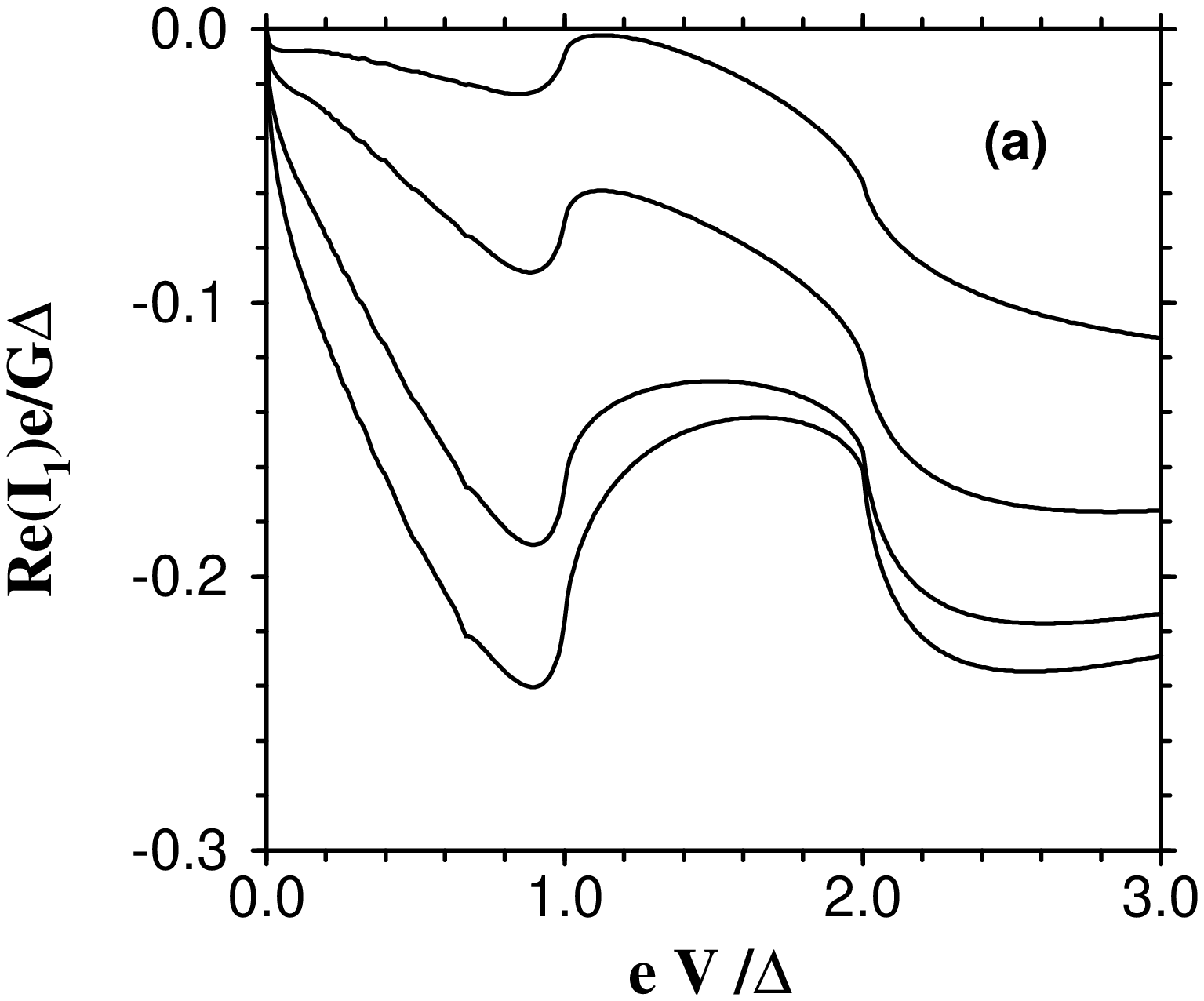,height=3.5in,angle=0}}
\centerline{
\psfig{file=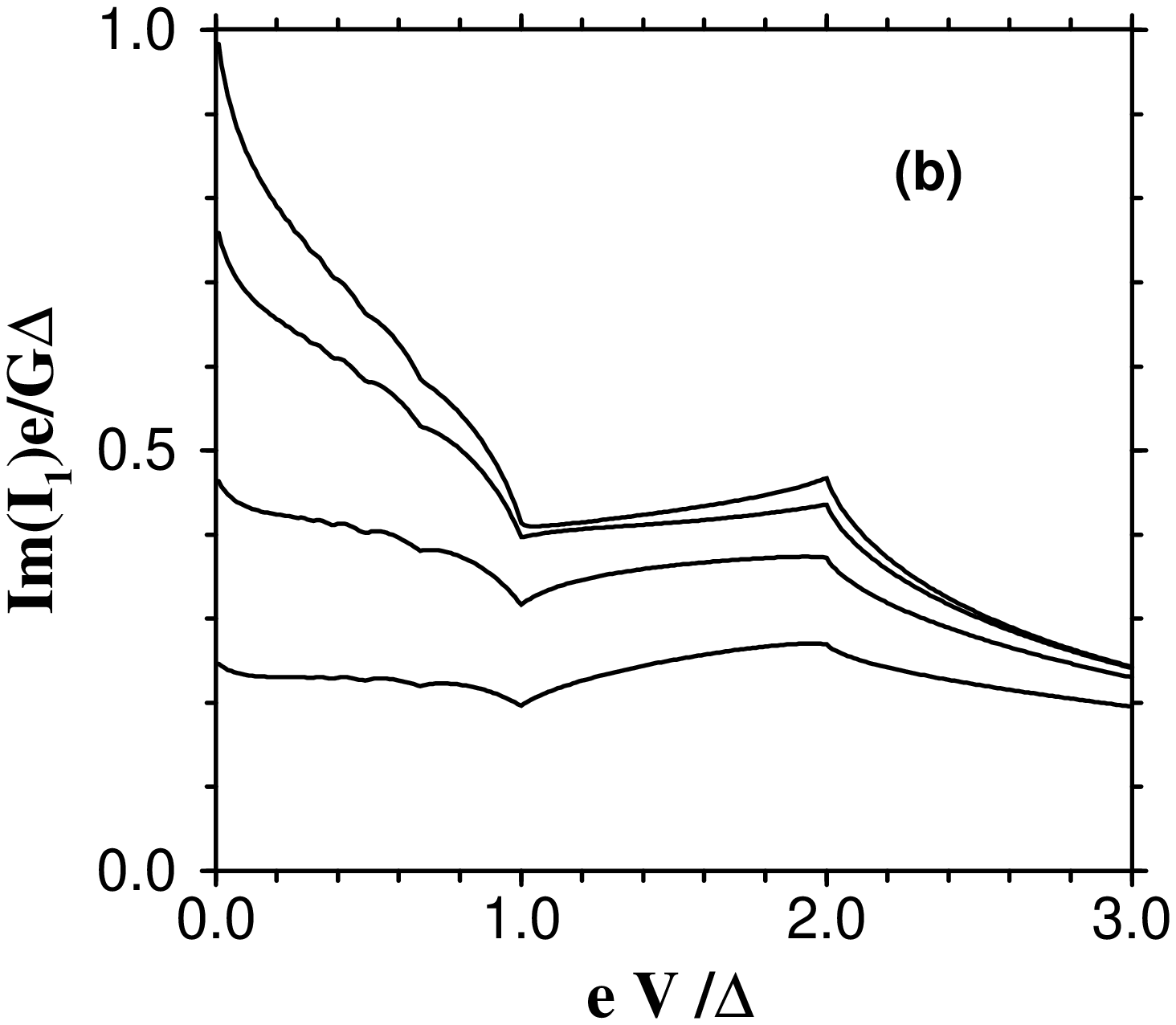,height=3.5in,angle=0}
}
\caption{
The cosine (a) and sine (b) part of the first Fourier
component of the ac current. From top to bottom (b) and 
from bottom to top (a) $T= 0, 1, 2, 3 \Delta$. 
The slope at small voltages diverges as $(\Delta/V)^{1/2}$.
See text for discussion about the high voltage behavior.}
\label{f3}
\end{figure}


\begin{references} 
\bibitem{kl} A.W. Kleinsasser, R.E. Miller, W.H. Mallison and
G.D. Arnold, Phys. Rev. Lett. {\bf 72}, 1738 (1994).
\bibitem{van} N. van der Post, E.T. Peters, I.K. Yanson and
J.M. van Ruitenbeek, Phys. Rev. Lett. {\bf 73}, 2611 (1994).
\bibitem{fr} A. Frydman and Z. Ovadyahu, Solid State Commun. 
{\bf 95}, 79 (1995).  
\bibitem{mur} L.C. Mur, C.J.P.M. Harmans, J.E. Mooij, J.F. Carlin,
 A. Rudra, and M. Ilegems, Phys. Rev.\ B {\bf 54}, {\bf R}2327 (1996).
\bibitem{ku} J. Kutchinsky, R. Taboryski, T. Clausen, C.B. Sorensen, 
A. Kristensen, P.E. Lindelof, J.B. Hansen, C.S. Jacobsen, and J.L. 
Skov, Phys.\ Rev.\ Lett. {\bf 78}, 931 (1997).
\bibitem{sac} E. Scheer, P. Joyez, D. Esteve, C. Urbina, and 
M.H. Devoret,  Phys.\ Rev.\ Lett. {\bf 78}, 3535 (1997). 
\bibitem{sh1} E.N. Bratus', V.S. Shumeiko, and G. Wendin,
Phys. Rev. Lett. {\bf 74}, 2110 (1995).
\bibitem{we1} D. Averin and A. Bardas, Phys. Rev. Lett. {\bf 75}, 1831 
(1995).
\bibitem{gun} U. Gunsenheimer and A.D Zaikin, Phys. Rev. \ B {\bf 50}, 
6317 (1994).
\bibitem{kbt} T.M. Klapwijk, G.E. Blonder, and M. Tinkham,
Physica {\bf B 109-110}, 1657 (1982).
\bibitem{zai1} A. Zaitzev, Sov.\ Phys.\ - JETP {\bf 51}, 111 (1980).  
\bibitem{fur} A. Furusaki and M. Tsukada, Sol.\ State Commun. {\bf 78}, 
299 (1991). 
\bibitem{been} C.W.J. Beenakker, Phys.\ Rev.\ Lett. {\bf 67}, 3836 
(1991).
\bibitem{bag} S. Datta, P.F. Bagwell, M.P. Anatram, Phys.\ Low-Dim.\ 
Struct. {\bf 3}, 1 (1996).  
\bibitem{oub} M.C. Koops, G.V. van Duyneveldt, and R. de Bruyn 
Ouboter, Phys.\ Rev.\ Lett. {\bf 77}, 2542 (1996).  
\bibitem{avz} S.N. Artemenko, A.F. Volkov, and A.V. Zaitsev, Sov.\ 
Phys.\ - JETP {\bf 49}, 924 (1979).  
\bibitem{naz} Yu. V. Nazarov, Phys. Rev. Lett. {\bf 73}, 134 (1994).
\bibitem{mr} A. Mart\'{i}n-Rodero, A. Levy Yeyati, and F.J. Garc\'{i}a-Vidal, 
Phys.\ Rev. B {\bf 53}, R8891 (1996). 
\bibitem{sh2} V.S. Shumeiko, E.N. Bratus' and G. Wendin, Low Temp.\ Phys.  
{\bf 23}, 181 (1997). 
\bibitem{ko2} I.O. Kulik and A.N. Omel'yanchuk, JETP Lett. {\bf 21}, 96 
(1975).
\bibitem{bau} K.M. Schep and G.E.W. Bauer, Phys.\ Rev.\ Lett. {\bf 78}, 
3015 (1997). 
\bibitem{al} L.G. Aslamazov and A.I. Larkin Sov.\ Phys.\ - JETP 
{\bf 43}, 698 (1976).  
\bibitem{kr} C. Nguyen, H. Kroemer, and E.L. Hu, Phys.\ Rev.\ Lett. 
{\bf 69}, 2847 (1992).  

\end{references}
\end{document}